\documentclass[9pt,twocolumn,twoside]{pnas-new}

\templatetype{pnasresearcharticle} 

\usepackage{epsfig}
\usepackage{graphicx,amssymb,amsmath,amsthm}
\usepackage{color}

\definecolor{red}{rgb}{1,0,0}
\definecolor{blue}{rgb}{0,0,1}
\definecolor{black}{rgb}{0,0,0}

\def\Ic{I}

\def\Lc{{\cal L}}

\def\Vc{{\cal V}}

\def\Dc{{\cal D}}
\def\Rc{{\cal R}}
\def\edot{\dot\epsilon}

\newcommand{\eq}[1]{\begin{align}#1\end{align}}

\newcommand{\ffrac}[2]{\mbox{$\frac{#1}{#2}$}}

\makeatletter
\def\pgfmath@parse@dimexpr@{%
  \edef\pgfmathresult{\pgfmath@tonumber{\pgfmath@dimen}}%
  \expandafter\pgfmath@stack@push@operand\expandafter{\pgfmathresult}%
  \pgfmath@parse@@operator%
  }
\makeatother

\usepackage{amsmath}
\newlength{\arrow}
\title{Friction law and hysteresis in granular materials}
\author[a]{E. DeGiuli} 
\author[a]{M. Wyart} 
\affil[a]{Institute of Physics, Ecole Polytechnique F\'ed\'erale de Lausanne (EPFL), CH-1015 Lausanne, Switzerland}

\leadauthor{DeGiuli} 

\significancestatement{The macroscopic friction of particulate materials often weakens as flow rate is increased.
This property is  responsible for important intermittent phenomena in geophysics, including earthquakes and landslides.
However, it is not understood  even in simple granular materials, where it is responsible for hysteresis of the angle of repose. Here we argue that for aerial granular flows, velocity-weakening is an emergent collective property induced by endogenous mechanical noise. Our theory rationalizes previously unexplained observations, and generates new, experimentally testable predictions. At a microscopic level it predicts that several microscopic quantities do not have a hard-particle limit, and characterizes their scaling behavior near jamming, which we confirm  numerically.}

\authorcontributions{Please provide details of author contributions here.}
\correspondingauthor{\textsuperscript{2}To whom correspondence should be addressed. E-mail: matthieu.wyart@epfl.ch}

\keywords{friction $|$ earthquakes $|$ granular media} 

\dates{This manuscript was compiled on \today}
\doi{\url{www.pnas.org/cgi/doi/10.1073/pnas.XXXXXXXXXX}}

\begin{abstract}
The macroscopic friction of particulate materials often weakens as the flow rate is increased, leading 
to potentially disastrous intermittent phenomena including earthquakes and landslides. 
We theoretically and numerically study this phenomenon in simple granular materials. 
We show  that velocity-weakening, corresponding to a non-monotonic behavior in the friction law $\mu(I)$, is present even if the dynamic and static microscopic friction coefficients  are identical, but disappears for softer particles. 
We argue that this instability is induced by endogenous acoustic noise, which tends to make contacts slide, leading to faster flow and increased noise. We show that soft spots, or excitable regions in the materials, correspond to rolling contacts that are about to slide, whose density is  described by a nontrivial exponent $\theta_s$. We build a microscopic theory for the non-monotonicity of $\mu(I)$, which also predicts the scaling behavior of acoustic noise, the fraction of sliding contacts $\chi$ and the sliding velocity, in terms of $\theta_s$. Surprisingly, these quantities have no limit when particles become infinitely hard, as confirmed numerically. Our analysis rationalizes previously unexplained observations and makes new experimentally testable predictions.
\end{abstract}

\begin{document}
\verticaladjustment{-2pt}

\maketitle
\thispagestyle{firststyle}
\ifthenelse{\boolean{shortarticle}}{\ifthenelse{\boolean{singlecolumn}}{\abscontentformatted}{\abscontent}}{}

\dropcap{A} central property of particulate materials is their macroscopic friction \cite{Forterre08,Schall10}. In velocity-strengthening materials, friction grows with the deformation rate and flow is stable. By contrast, velocity-weakening materials are susceptible to instabilities under loading, including stick-slip. This classification is believed to distinguish faults that are prone or not to earthquakes \cite{Scholz98}, which ultimately corresponds to the shear of a granular material made of rocks and debris -- the  gouge -- contained within the narrow fault between two tectonic plates. The Rice-Ruina rate-and-state model can describe both kinds of frictional behavior \cite{Baumberger06}. It is, however, a heuristic model, and building a microscopic theory to justify {\it a priori} which materials weaken or strengthen under flow remains a challenge.  To make progress, it is natural to consider well-controlled granular materials such as glass beads or sand, which have received considerable attention in the last decades.  In these systems an important result comes from dimensional analysis: assuming that grains are strictly hard and that this limit is not singular, the macroscopic friction or stress ratio $\mu\equiv \sigma/p$, where $\sigma$ is shear stress and $p$ is pressure, can be shown to depend on strain rate $\edot$ and pressure only via the dimensionless inertial number $\Ic \equiv \edot D \sqrt{\rho/p}$. Here  $D$ is mean grain diameter and $\rho$ is grain density \cite{MiDi04,Cruz05,Jop06}.  The constitutive relation $\mu(\Ic)$ has been extensively studied in the range $\Ic\geq 10^{-3}$ and is found to be a growing function of $\Ic$,  corresponding to a velocity-strengthening material. However, several recent experiments have reported non-monotonic behavior of $\mu(\Ic)$ for very small $\Ic$\cite{Dijksman11,Kuwano13,Wortel16}, corresponding to velocity weakening. There is currently no microscopic theory rationalizing these observations. 

\begin{figure}[t!]
\includegraphics[width=\columnwidth,clip]{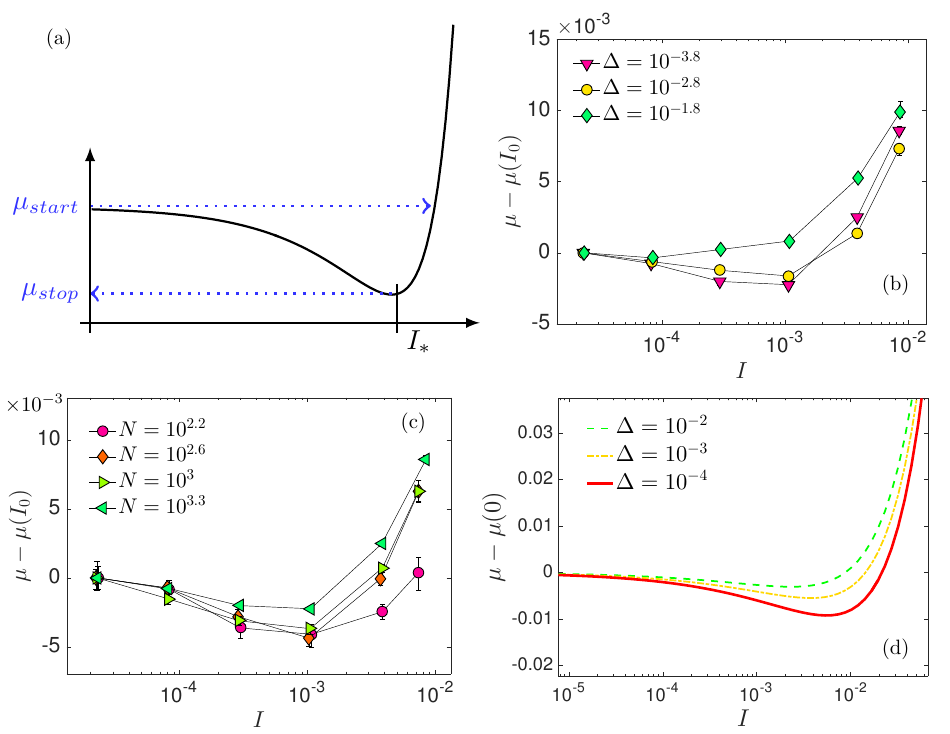}
\caption{ (a) Sketch illustrating that non-monotonic macroscopic friction leads to hysteresis. In a stress ramp from $\mu=0$, flow will begin at $\mu_{start}$, while in a stress decrease from large $\Ic$, flow will stop at $\mu_{stop}$ and a finite $\Ic$. (b,c) Numerical results for $\mu(\Ic,\Delta)-\mu(\Ic_0,\Delta)$ confirm the existence of a non-monotonic curve. In (b), $N=10^{3.3}$ and $\Delta$ is varied, while in (c) $\Delta \approx 10^{-4}$ and $N$ is varied. $\Ic_0=2 \times 10^{-5}$ is the smallest simulated $\Ic$.  (d) The qualitative behavior of $\mu(\Ic,\Delta)$ is reproduced by a phenomenological model (Eq. \ref{05}).
}\label{f0}
\end{figure}

As depicted in Fig.\ref{f0}a, the non-monotonic behavior of $\mu(\Ic)$ must lead to hysteresis effects  when $\mu$ is repeatedly cycled around its quasi-static value. Such hysteresis is well-known to characterize the jamming transition of granular materials. Indeed, the maximum angle of repose of a granular layer $\theta_{start}$ (corresponding to a macroscopic friction $\mu_{start}=\tan(\theta_{start})$) is larger than the angle $\theta_{stop}$ where avalanches stop (corresponding to $\mu_{stop}=\tan(\theta_{stop})$) \cite{Andreotti13}.  (Note that $\theta_{start}$ is measured by increasing the angle from a configuration that had stopped flowing at $\theta_{stop}$, otherwise  it is not well-defined and depends on system preparation.) Several properties of this hysteresis should be explained by a microscopic theory:  (i) externally applied vibrations \cite{Dijksman11, Johnson08, Lastakowski15,Wortel16} can eliminate hysteresis,  if the vibration velocity passes a threshold amplitude  \cite{Lastakowski15}.  
(ii) Hysteresis appears to become very small if inertial effects are negligible \cite{Dupont03}. 
(iii) There is no evidence of velocity weakening for frictionless particles \cite{Peyneau08}, as  confirmed below \footnote[1]{This statement does not contradict the observation that the macroscopic friction at which yielding occurs  depends on  preparation \cite{Xu06}. However, true velocity weakening in steady state without friction has only been observed for discontinuous, unphysical dissipation mechanisms \cite{Vaagberg17}.}.

In this letter we build a microscopic theory that explains these observations, and justify why for hard enough frictional particles $\mu(\Ic)$  is non-monotonic at small inertial numbers.  Our results, which we test using the discrete element method, hold even if the friction between two particles is assumed to be a simple Coulomb law with identical static and dynamic friction coefficients. This demonstrates that hysteresis emerges as a collective effect, even when absent at the contact level. Our approach is based on two previous fundamental insights. 
First, acoustic emissions generated by particle collisions during flow can fluidize the granular material, reducing dissipation. This idea was proposed by Melosh \cite{melosh79,Melosh96,Lucas14}  to explain why the macroscopic friction $\mu$ can be reduced by a factor of order $10$ from its static value during earthquakes. However, the quantitative treatment of this effect has been criticized \cite{Sornette00}, and the role of self-fluidization in regular granular materials is unclear, although self-fluidization was  experimentally shown to affect the density of the flowing material \cite{Elst12}. 

To quantify this role, we will use a second, recent insight: elementary excitations are a key feature of amorphous solids, in particular their density affects a material's stability and its susceptibility to plastic flow \cite{DeGiuli15a,Muller14,Lin14}. 
These excitations correspond to soft spots -- equivalent to dislocations in metals --  where local rearrangements can be triggered if the stress is increased, or if mechanical noise is present. For hard frictionless particles, excitations correspond to contacts carrying a small force, which are thus susceptible to open \cite{Wyart12,Lerner13a,DeGiuli15a}. 
 Here we will argue that for frictional granular materials, another kind of excitation appears, associated with contacts that are very close to sliding. Denoting by $\delta f$ the distance to sliding of a contact, we find that the distribution of excitations follows $P(\delta f)\sim \delta f^{\theta_s}$ with $\theta_s\approx -1/3$.  From this knowledge, we can compute the effect of mechanical noise on the structure, which in turn affects the noise itself. Ultimately this leads to an  experimentally testable microscopic theory of velocity weakening and mechanical noise. In addition, several non-trivial exponents are predicted that characterize noise and other microscopic quantities, which we find to be in good agreement with our numerical results.

\section{Numerical evidence for non-monotonic flow curves \label{s1}}

The macroscopic friction $\mu(I)$ has been studied extensively for inertial numbers $I\geq 10^{-3}$ \cite{MiDi04,Cruz05,Jop06,Azema14}. In the range $0.1\geq I\geq 10^{-3}$ it  follows
\eq{ \label{04}
\mu(\Ic) \approx \mu_c + c_1 \Ic^{\alpha_\mu}.
}
This velocity-strengthening behavior is caused by an increase in the dissipation induced by collisions as $I$ increases \cite{DeGiuli16}. There is, however, no microscopic theory for the value of $\alpha_\mu$, except for frictionless particles \cite{DeGiuli15a}.  For frictional particles one observes $\alpha_\mu \approx 0.85$ \cite{Peyneau09}, often approximated as $\alpha_{\mu}=1$.

Here we seek to numerically study if $\mu(I)$ becomes non-monotonic in the quasi-static regime $I\leq 10^{-3}$. However, if  velocity-weakening is indeed present, then homogeneous flow is unstable, and shear bands or chaotic behavior are expected. Experimental reports of velocity-weakening use a set-up with stress gradients \cite{Dijksman11,Kuwano13,Wortel16}, which has been argued to stabilize flow,  but makes quantitative analysis  challenging. Here instead we use the fact that  shear bands have a minimal thickness (typically of order ten grain diameters wide), and that flow can be stabilized by considering small systems.  Then, if $\mu$ has an instability, the non-monotonicity will be most pronounced for small $N$, and decay as larger systems are considered. 

In practice, our numerical results are created with the standard Discrete Element Method. We work in spatial dimension $d=2$ (our theoretical conclusions do not depend on $d$). The frictional disks follow static Coulomb friction: at each contact, the transverse component 
$f_T$ and normal component $f_N$ must satisfy $|f_T| \leq \mu_p f_N$, where we take the Coulomb friction coefficient to be $\mu_p=0.3$. This coefficient is the same for static and dynamic motion. Systems are simply sheared by horizontal motion of walls at the top and bottom domain edges, on a periodic domain. The confining pressure is controlled by fixed vertical forces exerted upon the walls. Particles have linear elastic-dashpot interactions, modeling a finite restitution coefficient $e<1$ and a particle stiffness $k$. $\Delta=p/k$, which  characterizes the typical contact deflection  relative  to the grain diameter, is varied from $\Delta \approx 10^{-2}$ to $\Delta \approx 10^{-5}$. 


Our results for  $\mu(\Ic)$ are shown in Figs. \ref{f0}b,\ref{f0}c. A crucial finding is that  non-monotonicity is indeed observed,  but disappears if particles are too deformed (large $\Delta$). For $\Delta \approx 10^{-4}, 10^{-3}$ we find that the minimum occurs at $\Ic_* \approx 10^{-3}$, the same inertial number below which intermittency and large fluctuations were previously observed in simulations \cite{Cruz05}. We measure the amplitude of non-monotonicity as $\Delta \mu_{hyst} \equiv \mu(\Ic_0) - \mu(\Ic_*)$ where $I_0=2 \times 10^{-5}$ is the smallest inertial number we probe. As expected, $\Delta \mu_{hyst}$ increases  as smaller systems are considered, but $\Ic_*$ does not vary significantly, as shown in Fig.\ref{f0}c and SI.

By contrast, for frictionless particles velocity weakening is essentially absent, or at least much weaker than for frictional particles, as shown in Fig.A.1,A.2 of SI. 
 

The existence of a minimum at any finite $\Ic_*$ implies a dramatic behaviour as the shear stress  is increased from the solid phase, as illustrated in Fig. \ref{f0}a, in particular  flow is predicted to start at a finite inertial number $I\approx 2 \times 10^{-3}$.  Essential questions include what microscopic mechanism leads to a minimum of $\mu(I)$ at $\Ic_*$, and why does this instability  disappear when $\Delta$ increases?

\section{Mechanism for instability in granular media} 
We argue that these questions are naturally explained if one considers the role of the acoustic noise endogenously generated in flow, as measured for example in \cite{Elst12}. In dense flows of hard particles,  a network of contacts is formed that constrains motion: particles cannot overlap and cannot slide if the considered contact satisfies the Coulomb criterion. For infinitely hard particles,  the dynamics only occurs along floppy modes for which these constraints are satisfied \cite{Lerner12a}. For any finite $\Delta$, there will exist small-amplitude motion orthogonal to the floppy modes, which corresponds to vibrations of the contact network. The formation of new contacts through collisions pumps energy into these vibrations, which eventually decays due to grain visco-elasticity.  Henceforth we call this vibrational energy per particle ``mechanical noise'' and denote it $E_{noise}$. If this noise is not negligible with respect to the characteristic potential energy in the contact, $E_p \propto \Delta^2$, it will affect the contact network and ``lubricate'' it \cite{Elst12,Leopoldes13}. In turn, this lubrication will facilitate particle motion, leading to faster flow, stronger collisions and larger noise. Our contention is that this positive feedback is responsible for instability in the flow curve and its associated hysteresis. This view naturally explains (i) why increasing particle deformability diminishes hysteresis, since it increases the potential energy which makes contacts less sensitive to noise, (ii) why immersing the grains in a viscous fluid reduces hysteresis, since it damps vibrations faster and (iii) why externally applied vibrations can eliminate hysteresis, since if the external noise is much larger than the endogenous noise, the positive feedback becomes negligible. 

\begin{figure}[t!]
\includegraphics[width=\columnwidth,clip]{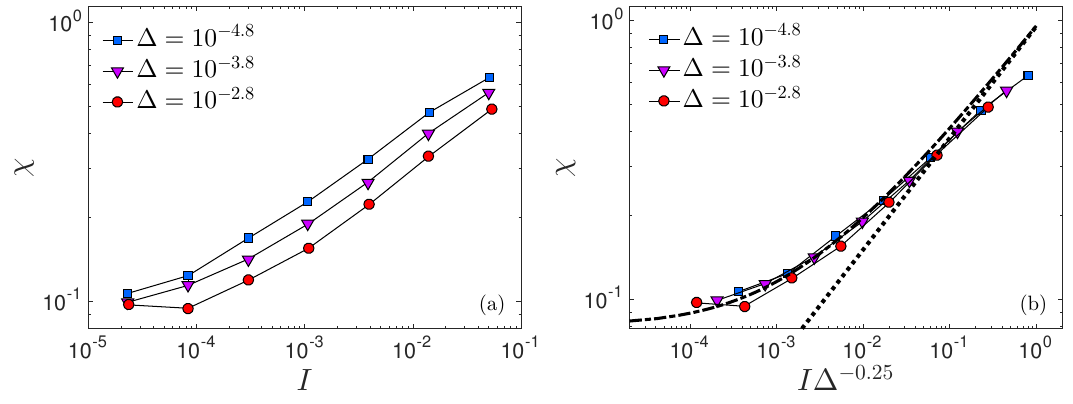}
\caption{ (a) The density of sliding contacts, $\chi$, depends both on $\Ic$ and $\Delta$. (b) This dependence can be collapsed by plotting $\chi$ vs $\Ic/\Delta^{1/4}$. The dotted line shows Eq.\ref{8}. If finite size effects are taken into account ($\chi_c>0$), theory predicts the dash-dotted line (SI). Here $N=10^{3}$.
}\label{f3}
\end{figure} 

We now argue more specifically that the dominant ``lubricating'' effect of mechanical noise is to intermittently remove rolling constraints, thus increasing the fraction of sliding contacts $\chi$. In previous work, such an increase was indeed observed in response to a transient forcing \cite{Ferdowsi14}. It is important to note that, with respect to noise, rolling constraints and no-overlap constraints are qualitatively different. Noise can lead to fluctuations in the distance between two particles in contact, but will never allow them to eventually penetrate each other. By contrast, vibration leading to intermittent stick-slip motion at a contact yields net relative motion between particles. This is illustrated by an object sitting on an inclined plane, which will eventually slide down if the plane is vibrated sufficiently. In our approach, this distinction explains why frictionless particles do not appear to present  noticeable hysteresis, as shown in SI \cite{Peyneau08}. (For very soft and elastic particles, other mechanisms than the removal of constraints described here could lead to a finite amount of hysteresis \footnote[7]{For highly deformable particles flow occurs via a succession of saddle-node bifurcations in the energy landscape, called shear transformations. Mechanical noise can then speed up flow by allowing jumps over barriers of potential energy, before these are destroyed by shear. This mechanism can cause hysteresis for very weak damping $e\approx 1$, see  \cite{Karimi17} for a recent discussion. In our framework such effects can be accounted for by replacing $\chi$ by the relative mechanical energy ${\cal R}$ in Eq.\ref{05}. However for realistic restitution coefficient and particle hardness this effect appears negligible, since it should exist in frictionless systems in which velocity weakening is essentially absent.}).

If  $\chi$ is indeed controlled by mechanical noise, then it should be a function of {\it both} $I$ and $\Delta$.  This prediction is confirmed in Fig. \ref{f3}, which shows that $\chi$ is  a function of $\Ic/\Delta^{1/4}$. 
Intriguingly, this result implies that the hard sphere limit $\Delta \rightarrow 0$ is singular, at least for some microscopic properties. In the next section we will build a quantitative theory of mechanical noise and explain this scaling property.  First, we argue why this behavior of $\chi$ can indeed generate hysteresis.



\begin{figure}[t!]
\includegraphics[width=\columnwidth,clip]{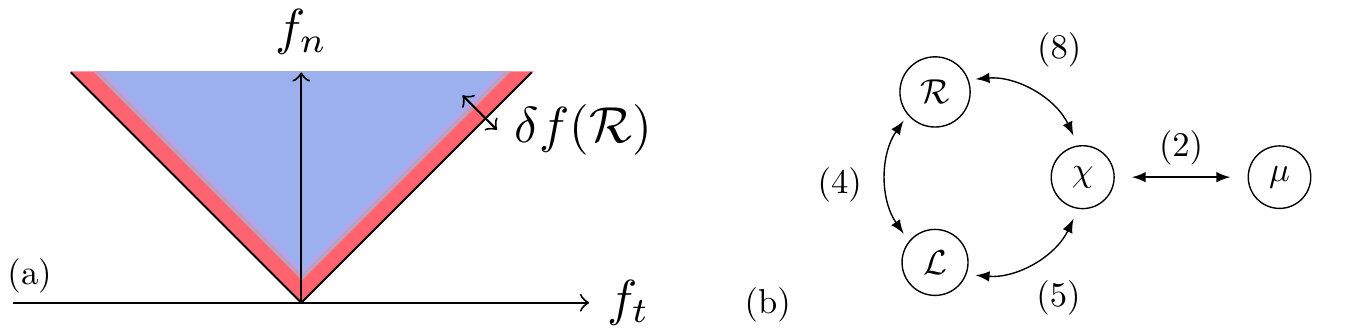}
\caption{ (a) Sketch of Coulomb cone. Mechanical noise of magnitude $\Rc$ induces force fluctuations of a size $\delta f(\Rc)$. Contacts within this distance of the edge of the Coulomb cone may be induced to slide by noise. (b) Schematic of logical relationships in theory between stress ratio $\mu$, density of sliding contacts $\chi$, dimensionless mechanical noise amplitude, $\Rc$, and dimensionless velocity scale, $\Lc$. Numbers denote relevant equations. 
}\label{f1}
\end{figure}

Consider the following Gedankenexperiment: the stress is increased in a granular solid. According to Eq.\ref{04}, flow will start when $\mu$ reaches $\mu_c$. Now, consider the same protocol, but with vibrations imposed on the sample. Because of this noise, some contacts that were close to the Coulomb cone, as pictured in Fig.\ref{f1}a, will intermittently slide, leading to an overall increase of the fraction of sliding contacts, $\chi$. Being less constrained, the material will be less stable and start flowing earlier, for some $\tilde {\mu_c} < \mu_c$. We write $\tilde{\mu}_{c} = \mu_c g(\chi)$, 
where $g$ is a decreasing function and $g(0) =1$. The same reduction of macroscopic static friction must apply  
when noise is endogenously generated in flow. This effect can be included in Eq.\ref{04}, which now becomes (taking $\alpha_\mu=1$ for simplicity):
\eq{ \label{05}
\mu(\Ic) \approx\tilde{\mu}_{c}+ c_1 \Ic= \mu_c +\mu_c (g(\chi)-1) + c_1 \Ic.
}
In this equation the last term is the usual velocity-strengthening effect due to the increased dissipation induced by collisions. 
The second-to-last term instead is velocity-weakening. To illustrate this point we consider a linear model $g(\chi) = 1 -b \chi$.
Qualitative results do not depend on the parameter $b$; here we choose $b=0.3$. Other parameters entering Eq.\ref{05} are fixed by previous observations to $c_1 \approx 1.4, \mu_c \approx 0.2$. Using fits to $\chi$ from Fig.\ref{f3} (described in SI),  Eq.\ref{05} predicts the curves shown in Fig. \ref{f0}d. We see that the  qualitative features of $\mu(I)$ are captured: $\mu$ has a minimum, whose depth, but not its location, depends strongly on $\Delta$. 

\section{Microscopic description of mechanical noise \label{s2}} 

Our approach to compute the mechanical noise and its effect on $\chi$ is illustrated in  Fig. \ref{f1}b. We proceed in three steps.
First, we express the mechanical noise in terms of two kinetic properties: the dimensionless velocity fluctuations $\Lc = V_r/(\edot D)$ (where  $V_r$ is the typical relative velocity between neighboring particles, $D$ the particle size and $\edot$ the strain rate) and $\epsilon_v$, the characteristic strain at which particles collide and change direction. Second, we use a previous argument based on energy balance relating $\chi$ to the normalized typical sliding velocity  ${\cal L}_T$. Finally,
 we compute the density of sliding contacts $\chi$ induced by mechanical noise. This requires knowing the density of rolling contacts at a small distance $\delta f$  to sliding. 
  From these three coupled equations quantities of interest are derived. 

Along the way we  make two approximations. First, we assume that the characteristic velocity ${\cal L}$ with which particles collide is identical to the characteristic sliding velocity ${\cal L}_T$. Assuming that there is only a single velocity scale in flow is essentially a mean-field approximation, known to be correct for frictionless particles \cite{DeGiuli15a} but not exact for frictional ones \cite{DeGiuli16}. 
Second, we assume that the vibrational noise is damped at a rate $1/\tau$ independently from the distance to jamming, which allows us to estimate the scaling of $E_{noise}$ directly in terms of the energy dissipated in collisions. The utility of these approximations is supported by the good comparison between predictions and numerics.

\subsection{Computing mechanical noise from kinetics} $E_{noise}$ characterizes the density of vibrational energy of the contact network. The power injected in vibrational modes  is supplied  by collisions at a rate $\Dc_{coll}$. It can be estimated as the product of the kinetic energy of a particle $\sim \Lc^2 \edot^2 m D^2$ times the collisional rate $\sim \edot/\epsilon_v$, leading to $\Dc_{coll} \sim \Lc^2  \edot^3 m D^2/\epsilon_v$ \cite{DeGiuli15a,DeGiuli16}. Introducing a time scale $\tau$  at which this vibrational energy is dissipated into heat, we get the estimate $E_{noise}=\tau\Dc_{coll} $. We make the simplifying assumption that $\tau$   does not depend on  $I$, characterizing the distance to jamming. Then dimensional analysis implies that $\tau= C_0 m/\eta_N$ where $\eta_N$ the  damping coefficient of the particle interaction. In our approximation $C_0$ is a constant, which we expect to be rather large \footnote[7]{In a network of interacting particles with linear dashpot interactions, the rate $1/\tau_\omega$ at which the vibrational energy in a mode of frequency $\omega$ decays follows $\tau_{\omega} \sim 1/\omega^2$. Thus the energy in low-frequency modes takes a much longer time to decay. It is well known that both amorphous solids \cite{DeGiuli15a} and contact networks in flow \cite{Lerner12a} display an abundance of low-frequency modes. Their frequency scale can be $\sim 10$ times smaller than the characteristic frequency of a single contact, suggesting the order of magnitude $C_0\sim 10^2$.}.
%
%
%
Normalizing $E_{noise}$ by the typical potential energy in a contact $E_{p} = \frac{1}{2} k \Delta^2 D^2$, we define their ratio
\eq{ \label{6}
\Rc \equiv \frac{E_{noise}}{E_{p}} \propto \frac{\Dc_{coll}}{\Delta^2} \frac{\tau}{k D^2}.
} 
Neglecting  pre-factors, we  obtain the scaling behavior:
\eq{ \label{7}
\Rc \sim Q \frac{\Lc^2 \Ic^3}{\Delta^{1/2} \epsilon_v} \sim Q \Lc^2 \left(\frac{\Ic}{\Delta^{1/4}}\right)^{2}, 
}
where  $Q\equiv \sqrt{mk/\eta_N^2}$ is the bare quality factor of the grains. We used the definition $\Ic \equiv \edot D \sqrt{\rho/p}$ as well as the previous result $\epsilon_v \propto I$ observed experimentally \cite{Menon97}, numerically \cite{DeGiuli15a,DeGiuli16} and justified theoretically in \cite{DeGiuli15a,DeGiuli16}.


\begin{figure}[t!]
\includegraphics[width=\columnwidth,clip]{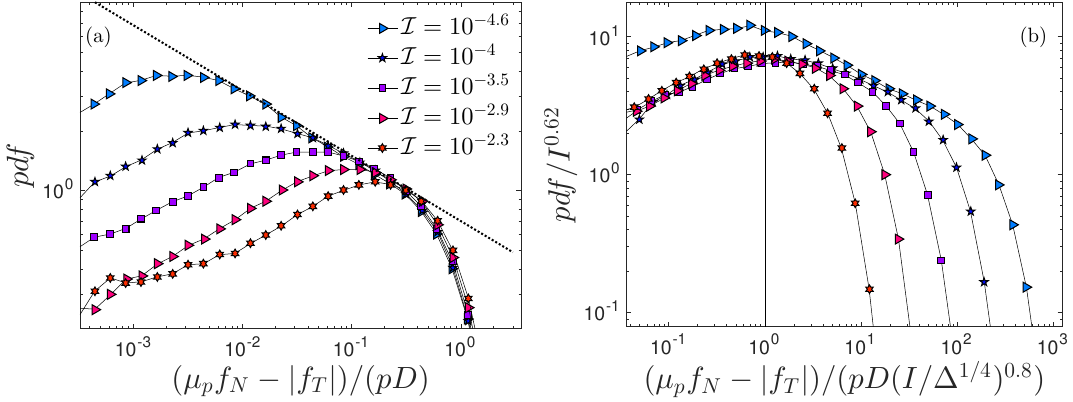}
\caption{ Probability density function (pdf) of distance to the edge of the Coulomb cone, rescaled by $p$, at $\Ic$ indicated in (b), for $\Delta \approx 10^{-5}$. (a) raw data, and (b) rescaled with $(\Ic/\Delta^{1/4})^{0.8}$. The dashed line shows a slope $-1/3$. 
}\label{f4}
\end{figure} 

\subsection{Constraint from energy balance} For realistic microscopic friction coefficient, it is found  in dense flows that most of the dissipation is induced by sliding at frictional contacts \cite{DeGiuli16} (the collisional dissipation $\Dc_{coll}$ is sub-dominant, but is the only one contributing to the mechanical noise). On average, the power dissipated in steady flow must balance the power injected by the shear stress; by a straightforward estimation of the sliding dissipation rate in terms of the typical sliding velocity \cite{DeGiuli16}, this implies  that \footnote{The power injected is $\Omega \sigma \edot$, where $\Omega$ is the system volume. At each sliding contact, the power dissipated is $f_T u_T$, where $f_T$ is the transverse force and $u_T\sim {\cal L}_T D \edot $ the transverse velocity. Equating the total power dissipated $\sim N \chi f_T u_T$ to $\Omega \sigma \edot$, we find $N D^3 \mu p \edot \sim N \chi \mu_p f_N \Lc_T \edot D$, or $\Lc_T \chi \sim  \mu / \mu_p $. For small $\Ic$, $\mu$ tends to a constant, so that $\Lc_T \chi \sim 1$.}:
\eq{ \label{8}
\Lc_T \propto 1/\chi.
}
where $\Lc_T$ is the characteristic dimensionless sliding velocity of sliding contacts. In what follows we assume $\Lc_T\sim {\cal L}$, which is an approximation \cite{DeGiuli16,Trulsson17}.

\subsection{Noise-induced sliding } For harmonic grains, the mechanical noise corresponds to a characteristic force scale 
\eq{ \label{9}
\delta {\tilde  f }= p D^{d-1} \Rc^{1/2}.
}
$\delta {\tilde  f }$ characterizes the fluctuations of forces at contacts. Such fluctuations will induce contacts {\it near} the edge of the Coulomb cone, shown in red in Fig.\ref{f1}, to slide intermittently. Thus, it is of crucial importance to determine the effect of noise on the density of contacts at small distances ${\tilde x}=f_N-|f_T|/\mu_p$ from the Coulomb cone. Fig.\ref{f4} shows the distribution $P(x)$ of the dimensionless quantity $x={\tilde x}/pD^{d-1}$. As $\Ic$ decreases, $P(x)$ develops a divergence $P(x)\sim x^{\theta_s }$ where $\theta_s \approx -1/3$.  At any time, we expect that about half of the contacts within a dimensionless distance $\delta f=\delta {\tilde  f }/p D^{d-1}$ of the Coulomb cone will be induced to slide, implying
\eq{ \label{10}
\chi - \chi_c \approx \frac{1}{2} \int_0^{\delta f} \; dx P(x),
}
where $\chi_c$ corresponds to the density of sliding contacts in the system when flow stops and noise is absent. As shown in SI, our numerics support that 
$\chi_c\rightarrow 0$ in the thermodynamic limit, but is finite for finite $N$. In what follows we therefore consider $\chi_c=0$, but our theory can readily be extended to the case $\chi_c>0$  to capture some finite size effects, see S.I.  
Inserting the scaling behavior $P(x) \sim x^{\theta_s}$ in Eq.\ref{10} leads to
\eq{ \label{11}
\chi  \propto \ffrac{1}{2} \Rc^{\frac{1+\theta_s}{2}}.
} 

\begin{figure}[t!]
\includegraphics[width=\columnwidth,clip]{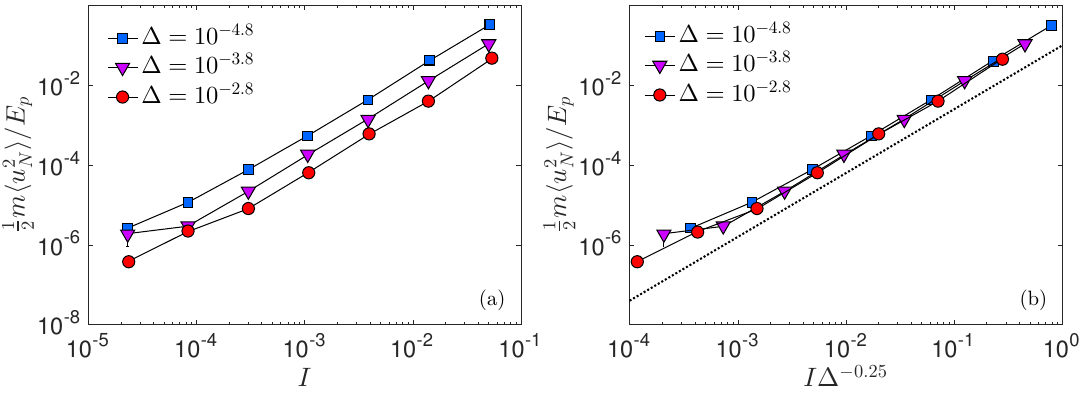}
\caption{ Ratio of estimated energy of mechanical noise $\frac{1}{2} m \langle u_N^2 \rangle$ to the potential energy scale $\frac{1}{2} k \Delta^2 D^2$. This is an approximation for $\Rc$ defined as  in Eq.\ref{6}, shown (a) vs $\Ic$, and (b) vs $\Ic/\Delta^{1/4}$. The dotted line in (b) has a slope 1.6. 
}\label{f5}
\end{figure}

\subsection{Results}
Combining Eqs.\ref{7},\ref{8},\ref{11} and keeping only on the dependence on $I$ and $\Delta$ leads to\footnote[2]{The $Q$ dependence is readily obtained in the Eqs.\ref{13},\ref{13*},\ref{13bis},\ref{13ter} by replacing $\Delta$ by $\Delta/Q^2$}:
%
\begin{eqnarray}
\label{13}
\chi &\sim& (\Ic/\Delta^{1/4})^{\alpha},\\
\label{13*} {\cal L}_T &\sim& (\Ic/\Delta^{1/4})^{-\alpha},\\
\label{13bis}{\cal R} &\sim& (\Ic/\Delta^{1/4})^{\gamma},\\
\label{13ter}\delta f &\sim&(\Ic/\Delta^{1/4})^{\beta}
\end{eqnarray}
where $\alpha = \frac{1+\theta_s}{2+\theta_s} \approx 2/5$,  $\gamma=\frac{2}{2+\theta_s}\approx1.2$ and $\beta=\frac{1}{2+\theta_s}\approx 0.6$. In Eq.\ref{13ter} the characteristic dimensionless fluctuation of forces is obtained from  Eq.\ref{9}, and follows $\delta f\sim {\cal R}^{1/2}$. These fluctuations are expected to smoothen the distribution of the distance to the Coulomb cone $P(x)$ for $x \ll \delta f$. $\delta f$ can thus be extracted numerically by locating where the power law $P(x)\sim x^{\theta_s}$ breaks down, as shown in Fig.\ref{f4}b.

\begin{figure}[t!]
\includegraphics[width=\columnwidth,clip]{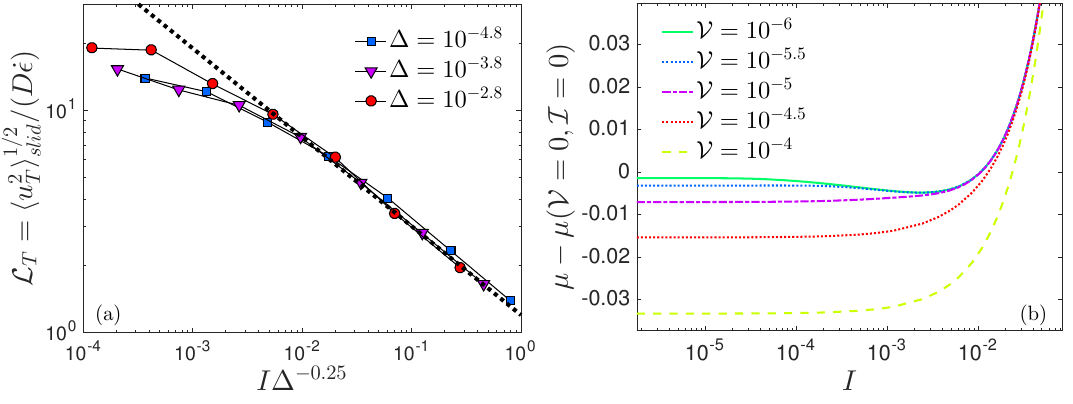}
\caption{ (a) The sliding velocity at contacts collapses when plotted vs $\Ic/\Delta^{1/4}$; dotted line shows $(\Ic/\Delta^{1/4})^{-2/5}$, predicted by theory. (b) Effect of external vibrations on rheology, for various dimensionless amplitude of external noise, $\Vc$, as predicted by the phenomenological model. 
}\label{f6}
\end{figure}

Eq.\ref{13} is tested in  Fig.\ref{f3}, Eq.\ref{13bis} in Fig.\ref{f5}, Eq.\ref{13ter} in Fig.\ref{f4} and Eq.\ref{13*} in Fig.\ref{f6}. 
Note that testing Eq.\ref{13bis} requires measurement of $E_{noise}$, which is difficult to do directly in the numerics, as it would require identification of all the vibrational modes of the contact network, at all instants of time.  To test Eq.\ref{13bis}  we measure a proxy for $E_{noise}$: the characteristic kinetic energy in the relative motion between particles $ m\langle u_N^2 \rangle/2$, where $u_N$ is the normal velocity at contact, as shown in Fig.\ref{f5}. It is  a lower-bound on $E_{noise}$  \footnote[7]{ Indeed for low-frequency vibrational modes, the kinetic energy is larger than the relative kinetic energy between neighboring particles.}.

Our prediction that the correct scaling variable for these macroscopic quantities  is $\Ic/\Delta^{1/4}$ is in perfect agreement with data. The predicted exponent $\alpha$ of Eq.\ref{13*} is in excellent  agreement with observations. This is also true for Eq.\ref{13} if finite size effects are included in the theory (see SI). Our measurements for the other exponents give $\gamma=1.6$ (Fig.\ref{f5}b) and $\beta=0.8$ (Fig.\ref{f6}b),  overall in good agreement with predictions, considering the approximations made. 


These results correspond to a detailed microscopic description of mechanical noise in granular materials. It is mean-field in character, but already appears to capture the essential aspects of the phenomenon. Obviously, the contact mechanics of real granular media can be much more complicated than the harmonic contact description discussed here \cite{Baumberger06,Jia11}.  For example, real grains in three dimensions are Hertzian, with an elastic potential $V(h) \propto E (h/D)^{5/2}$, where $E$ is the grain Young's modulus. For such grains, the typical dimensionless contact deflection $\Delta \propto (p/E)^{2/3}$ and the characteristic stiffness scales as $k \propto ED (p/E)^{1/6}$. This entails that $\delta f \propto E_{noise}^{3/5}$. Likewise, the damping coefficient $\eta_N$ may also depend on $\Delta$. Our framework can readily be extended to include  these different contact properties, which can slightly alter scaling exponents, but will not change the physical picture. 


Another  important possibility to consider is that the dynamic  friction coefficient is smaller than the static one. This will not affect the description of mechanical noise leading to Eqs.\ref{13},\ref{13*},\ref{13bis},\ref{13ter}. However, it will cause the macroscopic friction to decrease even more with $\chi$. In our phenomenological model, it corresponds to an increase in the coefficient $b$ appearing in Eq.\ref{05}, which will enhance the amplitude of hysteresis.

\section{External noise in granular flows} 
Applying external vibrations  is known experimentally to affect  rheological properties and hysteresis \cite{Dijksman11,Jia11,Melhus12,Leopoldes13,Lastakowski15,Wortel16}, as we now quantify. 
Even small typical noise can generate very rare events where noise is locally large, leading to rearrangements and creep flow. We do not seek to describe this effect here, which occurs at very small inertial number \cite{Wortel16}. Instead we focus on the dynamical range relevant to hysteresis and to our theory.  When a sheared material is subject to external noise, the dimensionless noise $\Rc$ will get an additional contribution. Let $E_e$ be the characteristic energy per grain from the external noise, and define $\Rc_{ext} \equiv E_e/E_p\sim(\Vc/\Delta)^2$, where $\Vc=\sqrt{E_e/(\frac{1}{2} k D^2)}$ and $\Rc_{tot} = \Rc + \Rc_{ext}$.
 Using Eqs.\ref{7},\ref{8} for $\Rc$, and Eq.\ref{11} with $\Rc_{tot}$ in place of $\Rc$, we find
\eq{  \label{14b}
\Rc_{tot} \sim Q \Rc_{tot}^{-1-\theta_s} \left(\frac{\Ic}{\Delta^{1/4}}\right)^{2} + \frac{\Vc^2}{\Delta^2}. 
}
Thus  external noise is relevant for $\Ic \lesssim \Ic^\dagger$, where $\Ic^\dagger~\sim~\Vc^{2+\theta_s}~\Delta^{-7/4-\theta_s}~Q^{-1/2}$. The results of Eq.\ref{14b} are shown in Fig. \ref{f6}b for a range of $\Vc$ (with $\theta_s=-1/3$).  There are two regimes: for weak noise,   $\mu$ remains non-monotonic, but  the value of $\mu_c$ is lowered. However, as noise is increased and $\Ic^\dagger > \Ic_*$, the non-monotonicity of $\mu$ disappears entirely. We thus predict that for sufficiently strong applied noise, flow becomes stable and hysteresis disappears. 


In the slider-block experiments of \cite{Lastakowski15},  a block is pulled over a granular bed in the presence of external vibrations of amplitude $A$ and frequency $\omega$. A transition from stick-slip to steady motion is observed, which does not depend on $A$ and $\omega$ independently, but only  on their product $A \omega$. This is expected from our analysis, since the external vibrational noise controlling the transition should be of order $E_e \sim \frac{1}{2} m (A \omega)^2$, indeed a function of $A \omega$ only.



Our framework can also predict the 
noise needed to  trigger flow in a static configuration with $\mu=\mu_c-\delta \mu$. Yielding will occur when the noise-induced reduction of static macroscopic friction is larger than $\delta \mu$, 
which according to Eqs.\ref{05},\ref{11} occurs  when $\delta \mu < b \mu_c \chi \sim  b \mu_c \big( \Vc/\Delta \big)^{1+\theta_s}$ or $\Vc > C \Delta (\delta \mu)^{1/(1+\theta_s)}$,
 where $C$ is a constant. This prediction is consistent with the earlier observation that the noise threshold increases with the confining pressure \cite{Johnson05, Xia13}. It would be interesting to further test the dependence of this threshold on $\delta \mu$.

\section{Conclusion \& Outlook}  We have shown that velocity-weakening  in dry granular flows is a collective phenomenon, which emerges even if absent at the contact level. We have explained this observation based on  a microscopic theory characterizing the  endogenous mechanical noise induced by collisions, as well as its effect on the structure. To understand the latter a characterization of elementary excitations on granular packings was performed, argued here to correspond to contacts that are rolling, but close to sliding. This framework rationalizes several experimental observations, including the factors governing the strength of hysteresis and the effects of exogenous noise. It also makes detailed scaling predictions on the noise level, the fraction of sliding contacts and the typical sliding velocity, in good agreement with  numerical observations. Several new predictions are made that can be  experimentally tested,  most obviously  the fact that velocity weakening should become very small or even vanish if sufficiently deformable particles are considered. This could be checked in an inclined plane experiment. From a theoretical perspective, an important question for the future is what determines the exponent $\theta_s$ characterizing excitations in frictional packings -- a question now well-understood for hard frictionless particles \cite{Wyart12, Charbonneau14}.

%
It is interesting to compare our results to Melosh's work on earthquakes \cite{melosh79,Melosh96}, where he proposed that acoustic emission fluidizes the fault gouge and  causes  the macroscopic friction to drop by a factor of order ten, as observed \cite{Di-Toro11}.  Melosh assumes that most of the energy dissipated goes into vibrations, whereas we find numerically in granular flows  that for inertial numbers of interest for velocity weakening  (say $I=10^{-3}$), this is only true for a few percent of the injected power \cite{DeGiuli16}. Likewise, in granular materials the hysteresis is a small effect, 10\% at most \cite{Forterre08}. 
Why such effects would be much larger in the context of a fault gouge is currently unclear, although it has been reported based on landslides that velocity weakening effects appear to grow with the length scale of the phenomenon considered \cite{Lucas14}.

Here we have focussed on $\mu$, but the particle volume fraction $\phi$ is also important in constitutive relations. Several works have observed that $\phi(\Ic)$ can become non-monotonic, corresponding to anomalous compaction \cite{Elst12,Kuwano13,Grob14} \footnote{In our numerical results we did not detect a noticeable non-monotonicity in $\phi(\Ic)$. As indicated by \cite{Elst12}, particle angularity may significantly enhance such anomalous compaction.}. In \cite{Elst12} this compaction was argued to arise from mechanical noise, and shown to diminish under the action of external vibrations, in a scenario qualitatively similar to our theory. Future work should carefully probe both $\mu$ and $\phi$ together to see if this link can be strengthened.


Finally, it would be desirable to extend  the present description to  over-damped non-Brownian dense suspensions.  Earlier experiments show that hysteresis increases with inertia, but do not rule out a finite amount of hysteresis in the viscous limit \cite{Dupont03,Clavaud17}. Over-damped simulations support that $\chi$ increases with  the viscous number \cite{Trulsson17} (playing a role similar to the inertial number), but very weakly. Following the discussion of Eq.\ref{05}, this behavior may be sufficient to lead to hysteresis. In that case, however, there is no microscopic theory for $\chi$ available.

\acknow{We thank M. Bouzid, E. Cl\'ement, Y. Forterre, X. Jia, B. Metzger, and V. Vidal for discussions. M.W. thanks the Swiss National Science Foundation for support under Grant No. 200021-165509 and the Simons Collaborative Grant ``Cracking the Glass Problem" \#454953. }

\showacknow 

 \pnasbreak 

\nocite{Kamrin14}
\bibliography{../../bib/Wyartbibnew}

\begin{thebibliography}{10}

\bibitem{Forterre08}
Forterre Y, Pouliquen O (2008) Flows of dense granular media.
\newblock {\em Annual Review of Fluid Mechanics} 40(1):1--24.

\bibitem{Schall10}
{Schall} P, {van Hecke} M (2010) {Shear Bands in Matter with Granularity}.
\newblock {\em Annual Review of Fluid Mechanics} 42:67--88.

\bibitem{Scholz98}
Scholz CH (1998) Earthquakes and friction laws.
\newblock {\em Nature} 391(6662):37--42 

\bibitem{Baumberger06}
Baumberger T, Caroli C (2006) Solid friction from stick--slip down to pinning
  and aging.
\newblock {\em Advances in Physics} 55(3-4):279--348 

\bibitem{MiDi04}
MiDi G (2004-08-01) On dense granular flows.
\newblock {\em The European Physical Journal E: Soft Matter and Biological
  Physics} 14(4):341--365.

\bibitem{Cruz05}
da~Cruz F, Emam S, Prochnow M, Roux JN, Chevoir Fmc (2005) Rheophysics of dense
  granular materials: Discrete simulation of plane shear flows.
\newblock {\em Phys. Rev. E} 72(2):021309.

\bibitem{Jop06}
Jop P, Forterre Y, Pouliquen O (2006) A constitutive law for dense granular
  flows.
\newblock {\em Nature} 441:727--730.

\bibitem{Dijksman11}
Dijksman JA, Wortel GH, van Dellen LT, Dauchot O, van Hecke M (2011) Jamming,
  yielding, and rheology of weakly vibrated granular media.
\newblock {\em Physical review letters} 107(10):108303.

\bibitem{Kuwano13}
Kuwano O, Ando R, Hatano T (2013) Crossover from negative to positive shear
  rate dependence in granular friction.
\newblock {\em Geophysical Research Letters} 40(7):1295--1299 

\bibitem{Wortel16}
Wortel G, Dauchot O, van Hecke M (2016) Criticality in vibrated frictional
  flows at a finite strain rate.
\newblock {\em Physical Review Letters} 117(19):198002.

\bibitem{Andreotti13}
Andreotti B, Forterre Y, Pouliquen O (2013) {\em Granular media: between fluid
  and solid}.
\newblock (Cambridge University Press).

\bibitem{Johnson08}
Johnson PA, Savage H, Knuth M, Gomberg J, Marone C (2008) Effects of acoustic
  waves on stick--slip in granular media and implications for earthquakes.
\newblock {\em Nature} 451(7174):57--60.

\bibitem{Lastakowski15}
Lastakowski H, G{\'e}minard JC, Vidal V (2015) Granular friction: Triggering
  large events with small vibrations.
\newblock {\em Scientific reports} 5.

\bibitem{Dupont03}
du~Pont SC, Gondret P, Perrin B, Rabaud M (2003) Granular avalanches in fluids.
\newblock {\em Physical review letters} 90(4):044301.

\bibitem{Peyneau08}
Peyneau PE, Roux JN (2008) Frictionless bead packs have macroscopic friction,
  but no dilatancy.
\newblock {\em Physical review E} 78(1):011307.

\bibitem{Xu06}
Xu N, O'Hern CS (2006) Measurements of the yield stress in frictionless
  granular systems.
\newblock {\em Phys. Rev. E} 73(6):061303.

\bibitem{Vaagberg17}
V{\aa}gberg D, Olsson P, Teitel S (2017) Shear banding, discontinuous shear
  thickening, and rheological phase transitions in athermally sheared
  frictionless disks.
\newblock {\em arXiv preprint arXiv:1703.01652}.

\bibitem{melosh79}
Melosh HJ (1979) Acoustic fluidization: a new geologic process?
\newblock {\em Journal of Geophysical Research: Solid Earth (1978--2012)}
  84(B13):7513--7520.

\bibitem{Melosh96}
Melosh H (1996) Dynamical weakening of faults by acoustic fluidization.
\newblock {\em Nature} 379:15.

\bibitem{Lucas14}
Lucas A, Mangeney A, Ampuero JP (2014) Frictional velocity-weakening in
  landslides on earth and on other planetary bodies.
\newblock {\em Nature communications} 5.

\bibitem{Sornette00}
Sornette D, Sornette A (2000) Acoustic fluidization for earthquakes?
\newblock {\em Bulletin of the Seismological Society of America}
  90(3):781--785.

\bibitem{Elst12}
van~der Elst NJ, Brodsky EE, Le~Bas P, Johnson PA (2012) Auto-acoustic
  compaction in steady shear flows: Experimental evidence for suppression of
  shear dilatancy by internal acoustic vibration.
\newblock {\em Journal of Geophysical Research: Solid Earth} 117.

\bibitem{DeGiuli15a}
DeGiuli E, D\"uring G, Lerner E, Wyart M (2015) Unified theory of inertial
  granular flows and non-brownian suspensions.
\newblock {\em Physical Review E} 91(6):062206.

\bibitem{Muller14}
M{\"u}ller M, Wyart M (2015) Marginal stability in structural, spin, and
  electron glasses.
\newblock {\em Annual Review of Condensed Matter Physics} 6(1):177--200.

\bibitem{Lin14}
Lin J, Lerner E, Rosso A, Wyart M (2014) Scaling description of the yielding
  transition in soft amorphous solids at zero temperature.
\newblock {\em Proceedings of the National Academy of Sciences}
  111(40):14382--14387.

\bibitem{Wyart12}
Wyart M (2012) Marginal stability constrains force and pair distributions at
  random close packing.
\newblock {\em Phys. Rev. Lett.} 109(12):125502.

\bibitem{Lerner13a}
Lerner E, During G, Wyart M (2013) Low-energy non-linear excitations in sphere
  packings.
\newblock {\em Soft Matter} 9(34):8252--8263.

\bibitem{Azema14}
Az{\'e}ma E, Radjai F (2014) Internal structure of inertial granular flows.
\newblock {\em Physical review letters} 112(7):078001.

\bibitem{DeGiuli16}
DeGiuli E, McElwaine J, Wyart M (2016) Phase diagram for inertial granular
  flows.
\newblock {\em Phys. Rev. E} 94:012904.

\bibitem{Peyneau09}
Peyneau PE ({2009}) Ph.D. thesis (Ecole des Ponts ParisTech).

\bibitem{Lerner12a}
Lerner E, D\"uring G, Wyart M (2012) A unified framework for non-brownian
  suspension flows and soft amorphous solids.
\newblock {\em Proceedings of the National Academy of Sciences}
  109(13):4798--4803.

\bibitem{Leopoldes13}
L{\'e}opold{\`e}s J, Conrad G, Jia X (2013) Onset of sliding in amorphous films
  triggered by high-frequency oscillatory shear.
\newblock {\em Physical review letters} 110(24):248301.

\bibitem{Ferdowsi14}
Ferdowsi B, Griffa M, Guyer R, Johnson P, Carmeliet J (2014) Effect of boundary
  vibration on the frictional behavior of a dense sheared granular layer.
\newblock {\em Acta Mechanica} 225(8):2227--2237 

\bibitem{Karimi17}
Karimi K, Ferrero EE, Barrat JL (2017) Inertia and universality of avalanche
  statistics: the case of slowly deformed amorphous solids.
\newblock {\em Physical Review E} 95(1):013003.

\bibitem{Menon97}
Menon N, Durian DJ (1997) Diffusing-wave spectroscopy of dynamics in a
  three-dimensional granular flow.
\newblock {\em Science} 275(5308):1920--1922 

\bibitem{Trulsson17}
Trulsson M, DeGiuli E, Wyart M (2017) Effect of friction on dense suspension
  flows of hard particles.
\newblock {\em Physical Review E} 95:012605.

\bibitem{Jia11}
Jia X, Brunet T, Laurent J (2011) Elastic weakening of a dense granular pack by
  acoustic fluidization: Slipping, compaction, and aging.
\newblock {\em Physical Review E} 84(2):020301.

\bibitem{Melhus12}
Melhus MF, Aranson IS (2012) Effect of vibration on solid-to-liquid transition
  in small granular systems under shear.
\newblock {\em Granular matter} 14(2):151--156 

\bibitem{Johnson05}
Johnson PA, Jia X (2005) Nonlinear dynamics, granular media and dynamic
  earthquake triggering.
\newblock {\em Nature} 437(7060):871--874 

\bibitem{Xia13}
Xia K, Huang S, Marone C (2013) Laboratory observation of acoustic fluidization
  in granular fault gouge and implications for dynamic weakening of earthquake
  faults.
\newblock {\em Geochemistry, Geophysics, Geosystems} 14(4):1012--1022 
  1525--2027.

\bibitem{Charbonneau14}
Charbonneau P, Kurchan J, Parisi G, Urbani P, Zamponi F (2014) Fractal free
  energy landscapes in structural glasses.
\newblock {\em Nature Communications} 5(3725).

\bibitem{Di-Toro11}
Di~Toro G, et~al. (2011) Fault lubrication during earthquakes.
\newblock {\em Nature} 471(7339):494--498 

\bibitem{Grob14}
Grob M, Heussinger C, Zippelius A (2014) Jamming of frictional particles: A
  nonequilibrium first-order phase transition.
\newblock {\em Physical Review E} 89(5):050201.

\bibitem{Clavaud17}
Clavaud C, B{\'e}rut A, Metzger B, Forterre Y (2017) Revealing the frictional
  transition in shear thickening suspensions.
\newblock {\em HAL preprint} pp. hal--01492671.

\bibitem{Kamrin14}
Kamrin K, Koval G (2014) Effect of particle surface friction on nonlocal
  constitutive behavior of flowing granular media.
\newblock {\em Computational Particle Mechanics} 1(2):169--176 

\end{thebibliography}

\;
\;
\;
\;
\;
\;
\;
\;
\;
\;

\pagebreak
\pnasbreak

\section{Supplementary Material}

\renewcommand\thefigure{A.\arabic{figure}} 

\renewcommand{\theequation}{A.\arabic{equation}}
\setcounter{figure}{0} 
\subsection{Friction and system-size dependence of hysteresis magnitude} 

\begin{figure*}[th]
\centering
\includegraphics[width=0.8\textwidth,clip]{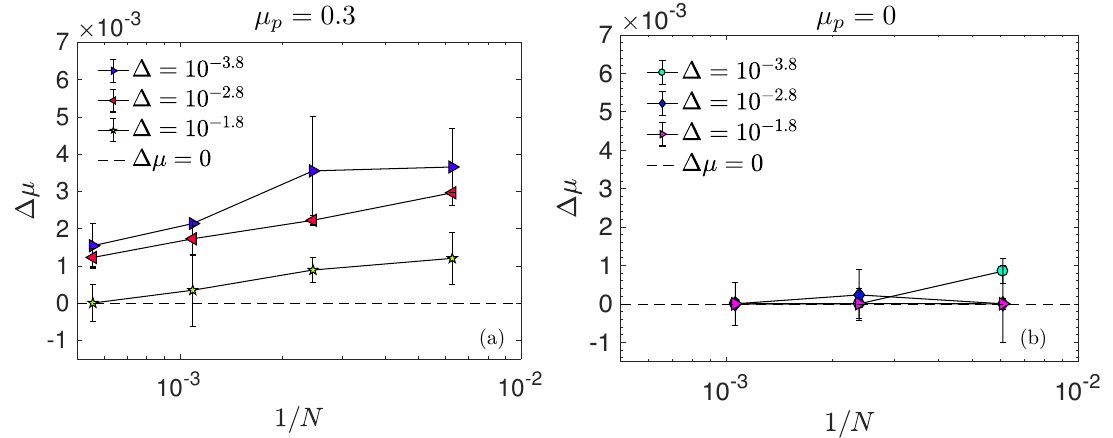}
\caption{ Hysteresis magnitude as a function of $N$ and $\Delta$, for (a) $\mu_p=0.3$, and (b) $\mu_p=0$. Hysteresis is essentially absent for frictionless particles.
}\label{fSI1}
\end{figure*}

As described in the main text, when smaller systems are considered, the magnitude of hysteresis is expected to increase. This can be inferred from Fig.1c in the main text. Defining $\Delta \mu_{hyst} \equiv \mu(\Ic = 2 \times 10^{-5}) - \mu(\Ic_*)$, we show directly the depth of the minimum for all data in Fig. \ref{fSI1}a. Error bars are  determined as follows: splitting the total strain into halves, for each run we compute $\Delta \mu_{hyst}$ in each half of strain. The error bar is the standard deviation of these values. 

\begin{figure*}[th]
\centering
\includegraphics[width=0.8\textwidth,clip]{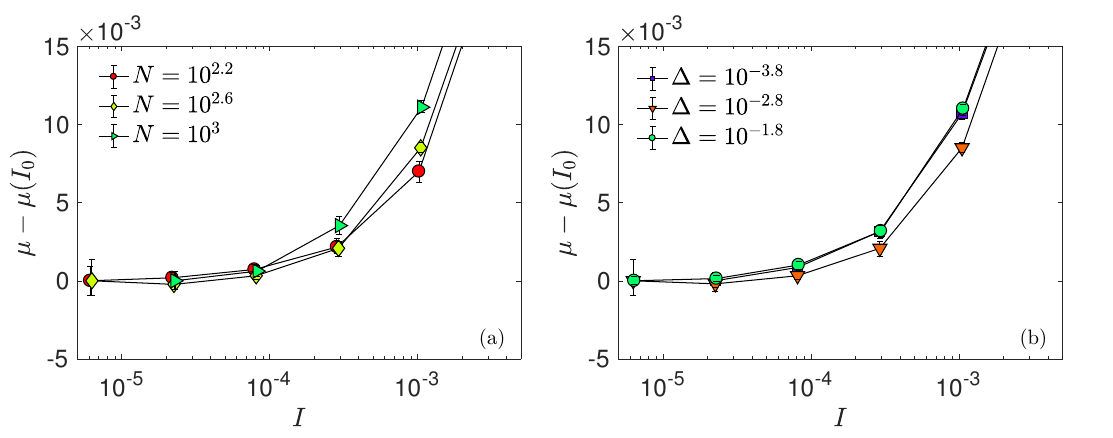}
\caption{ Curves of $\mu$ for frictionless systems, (a) at fixed $\Delta = 10^{-2.8}$ and (b) at fixed $N = 10^{2.6}$  
}\label{fSI2}
\end{figure*}

 We verified that for frictionless systems, hysteresis is absent or significantly smaller than for frictional particles. Following the same protocol as for the frictional systems, we computed $\mu(\Ic,\Delta,N)$ for $N=10^{2.2}, 10^{2.6}, 10^{3}$ and $\Delta \approx 10^{-4}, 10^{-3}, 10^{-2}$. In Fig. \ref{fSI2} we show example curves of $\mu$ at fixed $\Delta = 10^{-2.8}$ (Fig. \ref{fSI2}a) and at fixed $N = 10^{2.6}$ (Fig. \ref{fSI2}b). The horizontal and vertical range is equal to curves Fig. 1b, Fig. 1c in the main text. A dip is hardly distinguishable, but can be determined algorithmically in some cases, as for the frictional data. The resulting hysteresis magnitude is shown in Fig. \ref{fSI1}b and is seen to be negligible compared to the frictional case considered in the main text.

\subsection{Phenomenological Model Parameters} 
\begin{figure*}[bh!]
\centering
\includegraphics[width=0.8\textwidth,clip]{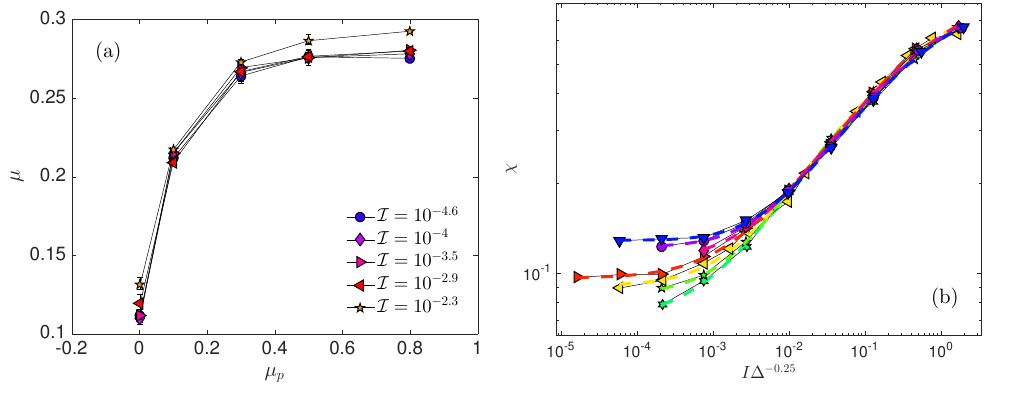}
\caption{ (a) Dependence of $\mu_c$ on the microscopic friction coefficient $\mu_p$. (b) Fits to Eq. \ref{chifit} for $\Delta = 10^{-3.8}$ and various $N$ (labels in Fig. \ref{fSI4}a).
}\label{fSI3}
\end{figure*}

The parameters $\mu_c$ and $c_1$ in the phenomenological model can be estimated from existing data \cite{Kamrin14,DeGiuli16}. Fig.\ref{fSI3}a shows $\mu$ for a range of $\mu_p$. For realistic $\mu_p$ in the range $0.2 < \mu_p < 0.8$, we find $0.23 < \mu < 0.27$. In the main text we take $\mu_c = 0.2$. 


We also use the linear form of $\mu(\Ic)$ at intermediate $\Ic$. Experiments and simulations are often fit to a form $\delta \mu(\Ic) = c_1 \Ic_1/(1+\Ic_1/\Ic)$, which behaves as $\delta \mu(\Ic) \approx c_1 \Ic$ for $\Ic$ in the dense regime, $\Ic \ll \Ic_1 \approx 0.3$ \cite{Kamrin14}. Here $c_1 \approx 1.4$ independently of $\mu_p$ \cite{Kamrin14}. 

Finally, to show the predictions of Eq.2, we use a functional form of $\chi(\Ic,\Delta)$ determined from data. As shown in Fig.\ref{fSI3} for $\Delta = 10^{-3.8}$ and various $N$, our data is very well fit by
\eq{ \label{chifit}
\chi = \frac{\sqrt{a_1 + a_2 (\Ic/\Delta^{1/4})^{2 a_3}}}{1 + a_4 (\Ic/\Delta^{1/4})^{a_3}}
}
The constant $a_1$ vanishes as $N\to \infty$; we use this large $N$ version in Fig. 1d in the main text.

\subsection{Finite-size effects in $\chi$} 
\begin{figure*}[bh!]
\centering
\includegraphics[width=0.8\textwidth,clip]{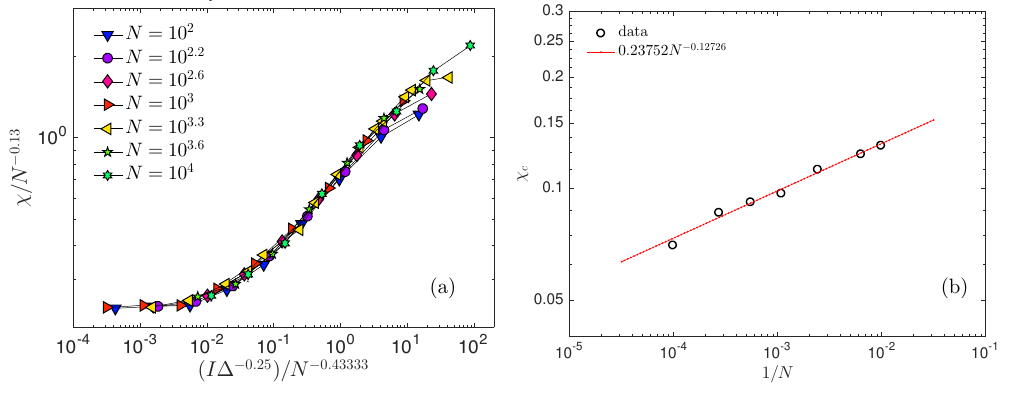}
\caption{ (a) Finite-size scaling for $\chi$ at $\Delta = 10^{-3.8}$. (b) Vanishing of $\chi_c = \chi(\Ic \to 0)$ as $N \to \infty$.
}\label{fSI4}
\end{figure*}

$\chi$ has a plateau value as $\Ic \to 0$ that depends on $N$. Using the fit form described above, we find that all data can be collapsed onto a master curve with finite-size scaling, as shown in Fig. \ref{fSI4}a for $\Delta = 10^{-3.8}$. The decay of $\chi_c = \chi(\Ic \to 0)$ with $N$ is shown in Fig.\ref{fSI4}b.

If the existence of $\chi_c$ is taken into account in the theory, this changes Eq.(8) to $\chi - \chi_c \propto \frac{1}{2} \Rc^{\frac{1+\theta_s}{2}}$, 
giving
\eq{ \label{12a}
\chi^2 (\chi-\chi_c)^{\frac{2}{1+\theta_s}} \propto Q \left(\frac{\Ic}{\Delta^{1/4}}\right)^{2}
}
Eq. \ref{12a} is a closed equation for $\chi(\Ic/\Delta^{1/4})$, and can be solved numerically. As in the main text, we predict that $\chi$ depends on $\Ic/\Delta^{1/4}$, as verified in Fig.2. The effect of $\chi_c$ in Eq. \ref{12a} is to saturate $\chi$ at small $\Ic/\Delta^{1/4}$, as shown by the dash-dotted line in Fig.2.

\end{document}